\documentclass[preprint,12pt]{elsarticle}

\usepackage{graphicx}
\usepackage{epsfig}

\usepackage{amssymb}

\journal{Physics Letters A}

\begin{document}
\begin{frontmatter}
\title{Open Quantum Walks on Graphs}

\author{S. Attal}
\address{Universit\'e de Lyon, Universit\'e de Lyon 1, C.N.R.S., Institut Camille Jordan, 21 av Claude Bernard, 69622 Villeubanne cedex, France}
\author{F. Petruccione}
\address{Quantum Research Group, School of Chemistry and Physics and National
Institute for Theoretical Physics, University of KwaZulu-Natal,
Durban, 4001, South Africa}
\author{I. Sinayskiy \corref{cor1}}
\address{Quantum Research Group, School of Chemistry and Physics and National
Institute for Theoretical Physics, University of KwaZulu-Natal,
Durban, 4001, South Africa}
\cortext[cor1]{Corresponding author. Tel./Fax: +27-(0)31-260-8133/8090\\ e-mail address: sinayskiy@ukzn.ac.za}

\begin{abstract}
Open quantum walks (OQW) are formulated as quantum Markov chains on graphs. It is shown that OQWs are a very useful tool for the formulation of dissipative quantum computing algorithms and for  dissipative quantum state preparation. In particular, single qubit gates and the CNOT-gate are implemented as OQWs on fully connected graphs. Also, dissipative quantum state preparation of arbitrary single qubit states and of all two-qubit Bell-states is demonstrated. Finally, the discrete time version of dissipative quantum computing is shown to be more efficient if formulated in the language of OQWs.
\end{abstract}

\begin{keyword}
Open quantum walk\sep dissipative quantum computing\sep dissipative state engineering 
\end{keyword}

\end{frontmatter}

\section{Introduction}
Recently, the experimental realization of a quantum computer has been the focus of extensive research \cite{QCR}. One of the main problems of the physical implementation of well known quantum algorithms \cite{QA} is the creation and manipulation of entanglement between qubits. Any physical system is subject to interaction with the environment, which inevitably leads to decoherence and dissipation \cite{toqs}. One way to compensate for this destructive environmental influence in the unitary implementations of the quantum algorithms is to introduce error-correcting codes \cite{QECC}. However, this approach treats the interaction with the environment as an effect the influence of which needs to be minimized.

A paradigm shift in looking for alternative strategies to realize quantum computers was induced with the theoretical prediction that dissipation can be used to create complex entangled states \cite{Diehl} and to perform universal quantum computation \cite{Vers}. This fundamental change is based on the assumption that one can manipulate the coupling of a system to an environment in such a way that the system is driven towards a thermal state, which is the solution of a particular quantum computing task \cite{Vers} or a target state in quantum state engineering \cite{Diehl,Vers}. The feasibility of this strategy was demonstrated by implementing dissipative quantum state engineering with ensembles of atoms \cite{Polzik} and trapped ions \cite{Blatt}. 

Quantum algorithms for universal quantum computing are conveniently formulated in the language of quantum walks \cite{uqwqc}. For example, a discrete time quantum walk implementation of the search algorithm on complex graphs has been shown to be more efficient than other known implementations of this quantum algorithm \cite{QS}. As for any other unitary implementation of quantum computing the efficiency of the quantum walk based realization decreases due to interaction with the environment. In view of the appeal of dissipative quantum computing it seems natural to formulate a dissipative version of the quantum walk so that algorithms for dissipative quantum computing and quantum state engineering are implemented efficiently. In other words, if dissipative quantum computing can make use of the interaction with the environment for performing universal quantum computation, can one introduce a framework which will use dissipative rather than unitary dynamics as a "driving force" of the quantum walk? 

During the last few years attempts were made to take into account dissipation and decoherence in the description of quantum walks \cite{ken1}. In these approaches decoherence is treated as an extra modification of the unitary quantum walk scheme, the effect of which needs to be minimized and eliminated. In fact, the general framework of quantum stochastic walks was proposed \cite{qsw}, which incorporates unitary and non-unitary effects of the quantum Markovian dynamics. In particular, by adding extra decoherence in experimental realizations of quantum walks, the transition from quantum to classical random walks was observed \cite{exp}.

Recently, a formalism for discrete time open quantum walks (OQW), which is exclusively based on the non-unitary dynamics induced by the environment was introduced \cite{APSS}. OQWs are formulated in the language of quantum Markov chains \cite{gudder} and rest upon the implementation of appropriate completely positive maps \cite{toqs, kraus}.

\section{Formalism}

To review briefly the formalism of OQWs, we consider a random walk on a set of nodes or vertices $\cal{V}$ with oriented edges $\{(i,j)\,;\ i,j\in\cal{V}\}$ as illustrated in Fig. 1(a).  The number of nodes is considered to be finite or countable infinite. The space of states corresponding to the dynamics on the graph specified by the set of nodes $\cal{V}$ will be denoted by  $\cal{K}=\mathbb{C}^\mathcal{V}$ and has an orthonormal basis ${(| i\rangle)}_{i\in\cal{V}}$. The internal degrees of freedom of the quantum walker, e.g. the spin or $n$-energy levels, will be described by a separable Hilbert space $\cal{H}$ attached to each node of the graph. More concisely, any state of the quantum walker will be described on the direct product of the Hilbert spaces $\cal{H}\otimes \cal{K}$. 
\begin{figure}
\begin{center}
\includegraphics[width= 0.5\linewidth]{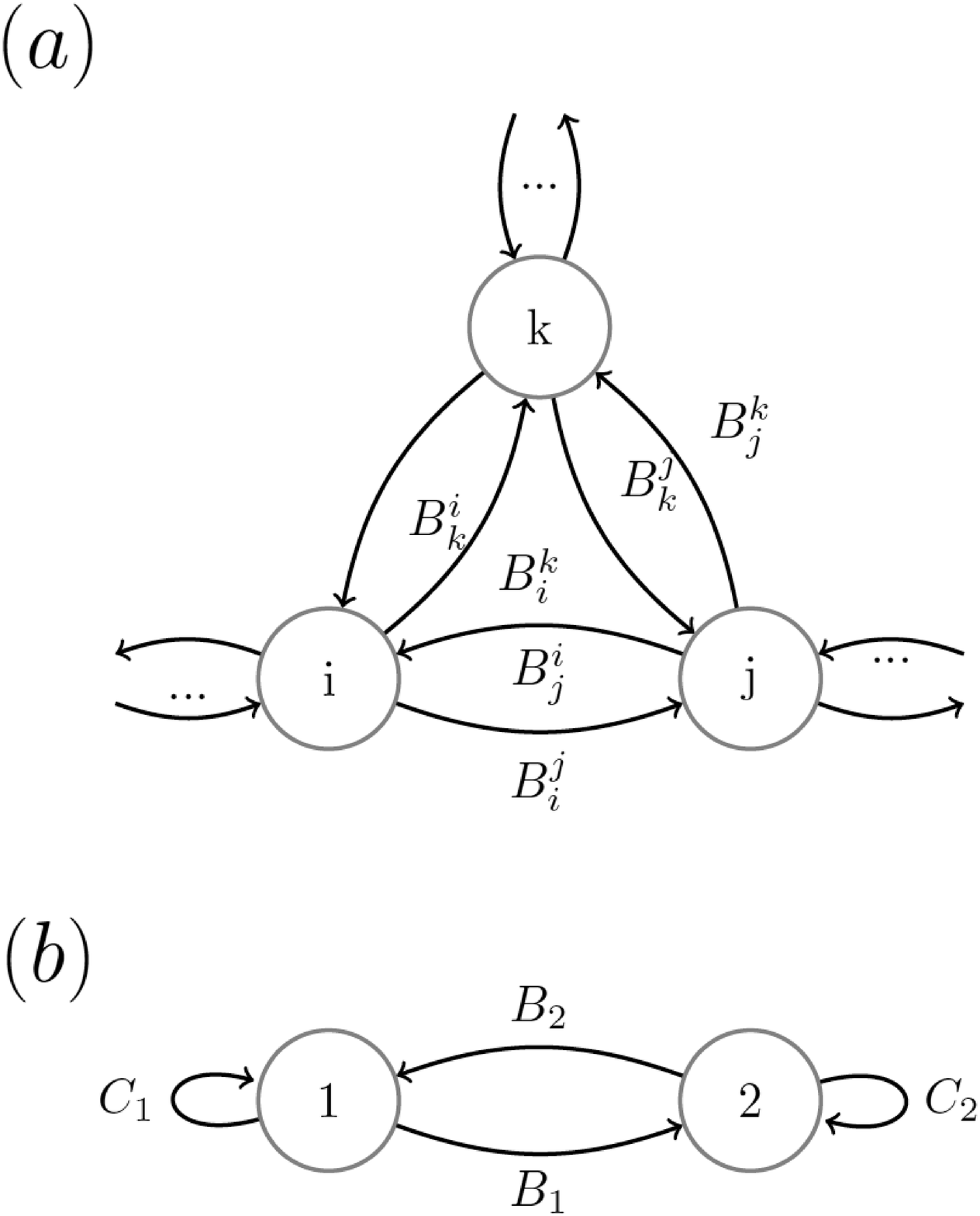} 
\caption{Schematic illustration of the formalism of the Open Quantum Walk. $(a)$ The walk is realized on a graph with a set of vertices denoted by $i,j,k\in\cal{V}$.  The operators $B_i^j$ describe transitions in the internal degree of freedom of the ``walker" jumping from node $(i)$ to node $(j)$. $(b)$ The simplest non-trivial example of the OQW on the finite graph is a walk on a two-node network. In this case the walk is performed using four operators $M_i^j (i,j=1,2)$. In particular, the transitions between node $1$ and node $2$ are induced by the operators $M_1^2=B_1\otimes|1\rangle\langle 2|$ and $M_2^1=B_2\otimes|2\rangle\langle 1|$; the operators describing changes in internal degrees of freedom of the "walker", if the "walker" does not jump, are $M_i^i=C_i\otimes|i\rangle\langle i|,(i=1,2).$}
\end{center}
\end{figure}

To describe the dynamics of the quantum walker, for each edge 
$(i,j)$ we introduce  a bounded operator $B^i_j\in\cal{H}$. This operator describes the change in the internal degree of freedom of the walker due to the shift from node $j$ to node $i$. By imposing for each $j$ that, 
\begin{equation}\label{eq1}
\sum_i {B^i_j}^\dag B^i_j= I,
\end{equation}
we make sure, that for each vertex of the graph $j\in\mathcal{V}$ there is a corresponding completely positive map on the positive operators of $\mathcal{H}$: 
$\mathcal{M}_j(\tau)=\sum_i B^i_j \tau {B^i_j}^\dag.$
Since the operators $B^i_j$ act only on $\mathcal{H}$ and do not perform transitions from node $j$ to node $i$, an operator $M^i_j\in\mathcal{H}\otimes\mathcal{K}$ can be introduced in the following form $M^i_j=B^i_j\otimes | i\rangle\langle j|.$
It is clear that, if the condition expressed in Eq. (\ref{eq1}) is satisfied, then $\sum_{i,j} {M^i_j}^\dag M^i_j=1$. This latter condition defines a completely positive map for  density matrices on $\mathcal{H}\otimes\mathcal{K}$, i.e.,
\begin{equation}\label{DQRW}
\mathcal{M}(\rho)=\sum_i\sum_j M^i_j\,\rho\, {M^i_j}^\dag.
\end{equation}
The above  map defines the discrete time \textit{open quantum walk} (OQW).
For an arbitrary initial state the density matrix $\sum_{i,j} \rho_{i,j}\otimes|i\rangle\langle j|$ will take a diagonal form after just one step of the open quantum walk. By the direct insertion of an arbitrary initial condition in Eq. (\ref{DQRW}) we get
\begin{eqnarray}
\nonumber
\mathcal{M}\left(\sum_{k,m} \rho_{k,m}\otimes|k\rangle\langle m|\right)&=&\sum_{i,j,k,m} B^i_j\otimes | i\rangle\langle j|\,\left(\rho_{k,m}\otimes|k\rangle\langle m|\right)\, {B^i_j}^\dag\otimes | j\rangle\langle i|\\\nonumber
&=&\sum_{i,j,k,m} B^i_j\rho_{k,m}{B^i_j}^\dag\otimes | i\rangle\langle i|\delta_{j,k}\delta_{j,m}\\
&=&\sum_i \left(\sum_j B^i_j\rho_{j,j}{B^i_j}^\dag\right)\otimes | i\rangle\langle i|.
\end{eqnarray}
Hence, we will assume that the initial state of the system has the form $\rho=\sum_i \rho_i\otimes |i\rangle\langle i|$, with $\sum_i \mathrm{Tr}[\rho_i]=1$.  
It is straightforward to give an explicit formula for the iteration of the OQW from step $n$ to step $n+1$: $\rho^{[n+1]}=\mathcal{M}(\rho^{[n]})=\sum_i \rho_i^{[n+1]}\otimes |i\rangle\langle i|$, 
where $\rho_i^{[n+1]}=\sum_j B^i_j \rho_j^{[n]}{B^i_j}^\dag$. This iteration formula gives a clear physical meaning to the mapping $\mathcal{M}$: the state of the system on  site $i$ is determined by the conditional shift from all connected sites $j$, which are defined by the explicit form of the operators $B_j^i$. Also, one can see that $\mathrm{Tr}[\rho^{[n+1]}]=\sum_i\mathrm{Tr}[\rho_{i}^{[n+1]}]=1$. Generic properties of OQWs have been discussed in \cite{APSS}.

As a first illustration of the application of the formalism of open quantum walks we consider the walk on a 2-node graph (see Fig. 1b). To be specific, the transition operators are $M_1^2=B_1\otimes|1\rangle\langle 2|$ and $M_2^1=B_2\otimes|2\rangle\langle 1|$ and the operators describing changes in internal degrees of freedom of the "walker", if the "walker" does not jump, are $M_i^i=C_i\otimes|i\rangle\langle i|,(i=1,2)$. In this case for each node $(i=1,2)$ we have:
\begin{equation}
B_i^{\dag}B_i+C_i^{\dag}C_i=I.
\end{equation}
The state $\rho^{[n]}$ of the walker after $n$ steps is given by,
\begin{equation}
\rho^{[n]}=\rho_1^{[n]}\otimes|1\rangle\langle 1|+\rho_2^{[n]}\otimes|2\rangle\langle 2|,
\end{equation}
where the particular form of the $\rho_i^{[n]}$ $(i=1,2)$ is found by recursion,
\begin{eqnarray}
\rho_1^{[n]}=C_1\rho_1^{[n-1]}C_1^{\dag}+B_2\rho_2^{[n-1]}B_2^{\dag},\\\nonumber
\rho_2^{[n]}=C_2\rho_1^{[n-1]}C_2^{\dag}+B_1\rho_2^{[n-1]}B_1^{\dag}.
\end{eqnarray}
\begin{figure}
\begin{center}
\includegraphics[width= 0.5\linewidth]{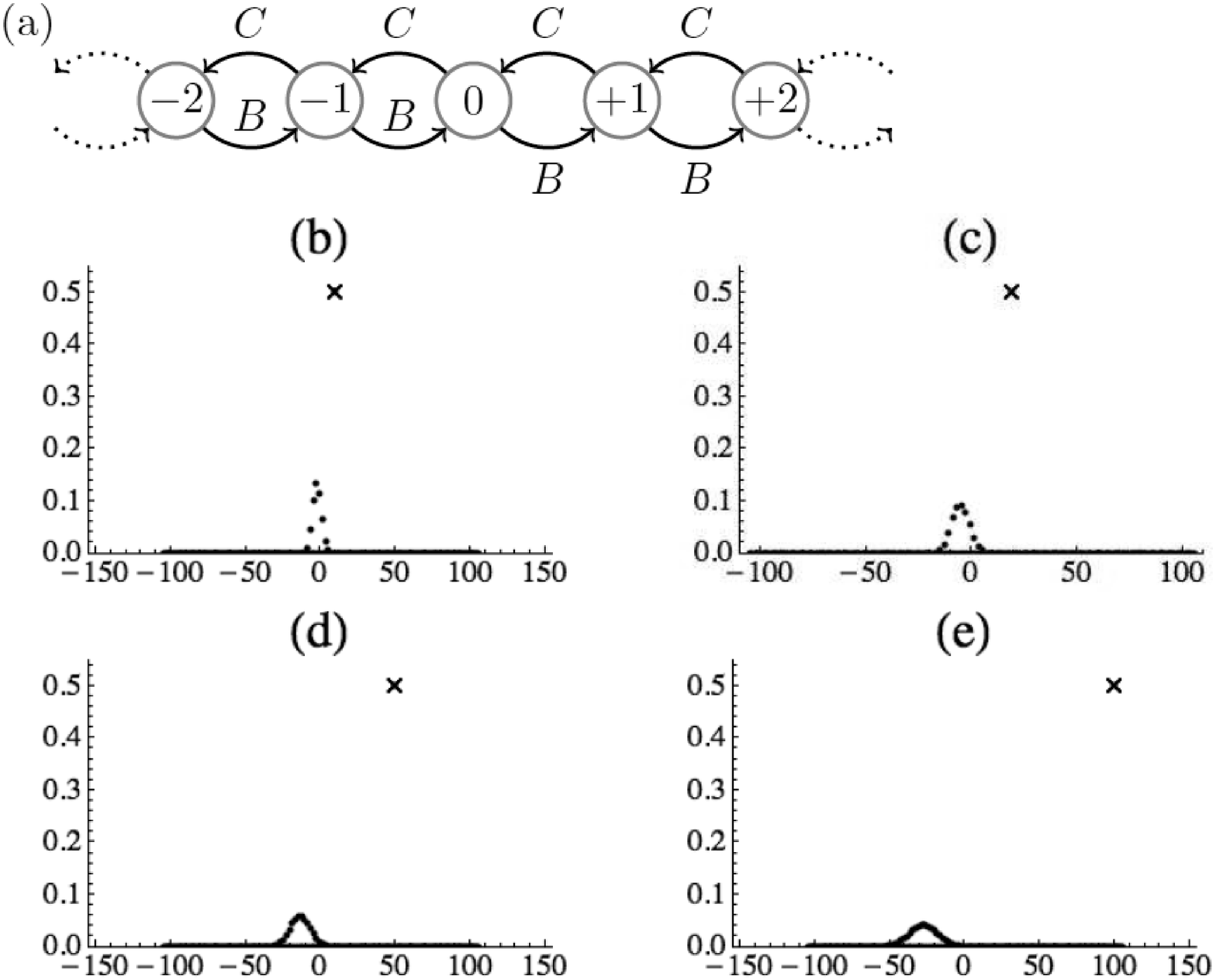} 
\caption{Open quantum walk on $\mathbb{Z}$. (a) A schematic representation of the  OQW on $\mathbb{Z}$: all transitions to the right are induced by the operator $B_i^{i+1}\equiv B$, while all transitions to the left are induced by the operator $B_i^{i-1}\equiv C$ (see Eq. (\ref{BCZ})); Figures (b)-(e) show the occupation probability distribution for the ``walker" with the initial state $I_{2}/2 \otimes |0\rangle\langle 0|$ and transition operators given by Eq. (\ref{BCZ}) after 10, 20, 50 and 100 steps, respectively. Two distinctive behaviors of the "walker" are observed: a Gaussian wave-packet moving slowly to the left (dots) and a deterministic
trapped state propagating to the right at a speed of 1 in units of cells per time step (cross).}
\end{center}
\label{fig:DQRWZ1}
\end{figure}

Next we consider an open quantum walk on the line.
The situation is depicted in Fig. \ref{fig:DQRWZ1} (a). For this simple one-dimensional walk, the only non-zero transitions out of cell $i$ are given by operators of the form $B_i^{i+1}\equiv B$ and $B_i^{i-1}\equiv C$. Obviously, the operators $B$ and $C$ satisfy the condition $B^\dag B+C^\dag C=I$, as imposed in Eq. (\ref{eq1}). Assuming the initial state of the system to be localized on site $0$, i.e., $\rho_0=\rho\otimes |0\rangle\langle 0|$,  after one step the system will jump to sites $\pm 1$ so that the new density matrix will be $\rho^{[1]}=B\rho B^\dag\otimes |1\rangle\langle 1|+C\rho C^\dag\otimes |-1\rangle\langle -1|$. The procedure can easily be iterated.  In Figs.  \ref{fig:DQRWZ1}(b)-(e) we show the probability to find a ``walker" on a particular lattice site for different numbers of steps. For this simulation we have chosen the transition matrices $B$ and $C$ as follows,
\begin{equation}\label{BCZ}
B=\sin{\theta}|-\rangle\langle-|+|+\rangle\langle+|,\quad C=\cos{\theta}|-\rangle\langle-|,
\end{equation}
and $\cos{\theta}=4/5$. One can see that already after 10 steps, there are two distinctive behaviors of the ``walker". The first is a Gaussian wave-packet moving slowly to the left and the second one is a completely deterministic trapped state propagating to the right at a speed of 1 in units of cells per time step. Interestingly, the state of the ``walker" in the ``soliton like" part is given by $\frac{1}{2}|+\rangle\langle+|$, while in the other, Gaussian part is given by the $p_n|-\rangle\langle-|$, where, of course $|+\rangle$ and $|+\rangle$ states are defined as $ |\pm\rangle=(|0\rangle\pm|1\rangle)/\sqrt{2}$ and $p_n$ is the probability to find  the ``walker" on the site $n$. Even this simple example demonstrates a remarkable dynamical richness of OQW. Further examples of OQW on $\mathbb{Z}$ which show a behavior distinct from the classical random walk and the unitary quantum walk can be found in \cite{APSS}.

\section{OQW for dissipative quantum computing and state engineering}

\begin{figure}
\begin{center}
\includegraphics[width=.9\textwidth]{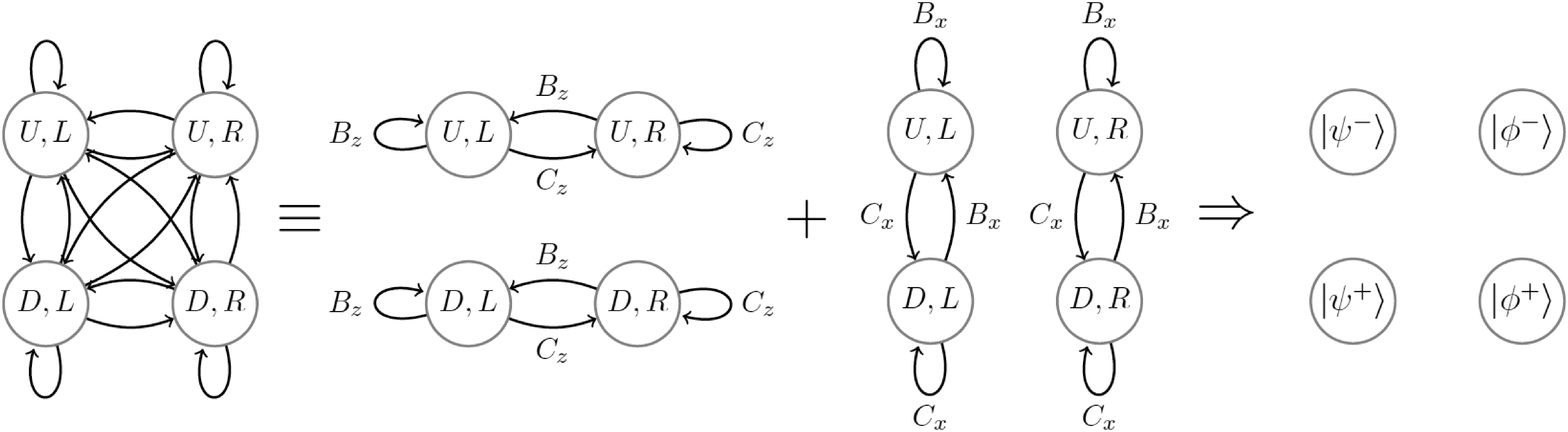}
\caption{OQW preparation of the Bell-states of two qubits. A generic OQW walk on a 4-node graph is decomposed in two walks on two-nodes graphs: ``left-right" and ``down-up" walks. For an initial unpolarized state of two-qubits on any nodes $\rho_0=\frac{1}{4}I_4\otimes|i\rangle\langle i|$ the corresponding Bell-pairs are obtained by measuring the position of the walkers after performing the OQW.}
\end{center}
\end{figure}

To motivate the potential of the suggested approach for the formulation of quantum algorithms for dissipative quantum computing and quantum state engineering we consider the example of an OQW on the 2-node fully connected graph in Fig. 1 (b). This example, will show that it is possible to implement all single-qubit gates and the CNOT-gate in the language of the suggested formalism. To be specific, in order to realize an X-gate with OQWs we prepare the system in some initial state $|\psi_0\rangle$ in node 1 and we will read the result of the computation in the node 2 (See Fig. 1(b)). If we choose $B_1=\sqrt{p}X$, $C_1=\sqrt{q}I_2$, $B_2=\sqrt{q}X$ and $C_2=\sqrt{p}I_2$, where $p$ and $q$ are positive constants such that $p+q=1$, then the stationary state of this walk will have the following form $\rho_{SS}=q|\psi_0\rangle\langle\psi_0|\otimes|1\rangle\langle 1|+pX|\psi_0\rangle\langle\psi_0|X\otimes|2\rangle\langle 2|$. Therefore, the OQW on this graph realizes the X-gate with probability $p$ upon read-out of the presence of the ``walker" in node 2. In a similar way, in order to implement the CNOT-gate we initially place two ``walkers" in node 1. We choose $B_1=\sqrt{p}U_{CNOT}$, $C_1=\sqrt{q}I_4$, $B_2=\sqrt{q}U_{CNOT}$ and $C_2=\sqrt{p}I_4$ and if the presence of both ``walkers" is measured in the node 2 then the OQW realized the CNOT-gate with probability $p$.

On the same 2-node network we can also implement dissipative state preparation (see Fig. 1(b)). To this end, we consider trivial transition matrices on the node 1, i.e., $B_1=C_1=I_2/\sqrt{2}$ and non-unitary transition matrices on the node 2, i.e., $B_2=\sqrt{p}|\psi^{(1)}\rangle\langle\psi^{(2)}|$ and $C_2=\sqrt{q}|\psi^{(2)}\rangle\langle\psi^{(2)}|+|\psi^{(1)}\rangle\langle\psi^{(1)}|$, where $p$ and $q$ are positive constants such that $p+q=1$, $|\psi^{(1)}\rangle=(\cos{\alpha},\sin{\alpha}e^{-i\beta})^\dag$ and $|\psi^{(2)}\rangle=(-\sin{\alpha},\cos{\alpha}e^{-i\beta})^\dag$. With this choice an arbitrary initial state will converge to a unique steady state, namely $\rho_f=|\psi^{(1)}\rangle\langle\psi^{(1)}|\otimes|2\rangle\langle2|$. The probability of detecting the system in the steady state after $2m$ steps of the walk is given by $P_{SS}\sim1-\rho^{(1)}_{22}(0)/{4^m}-(\rho^{(1)}_{11}(0)+\rho^{(2)}_{11}(0))[\mathrm{min}(1/4,q)]^m$, where $\rho^{(i)}_{jj}(0)$ are the elements of the initial density matrix of the system, $\rho^{(i)}_{jj}(0)=\langle i,\psi^{(j)}|\rho(0)|i,\psi^{(j)}\rangle$.

OQWs on more complex graphs allow the dissipative preparation of entangled multi-qubit states. With two ``walkers" on a 4-node network (see Fig. 3) we can prepare all two-qubits Bell-states. In this particular case, the OQW can be decomposed in a combination of two walks on two independent 2-node networks. The first ``walker" moves up and down, while the second one moves left and right (see Fig.3). We choose the transition operators to be $B_z=\frac{1}{2}(I-Z_1Z_2)$, $C_z=\frac{1}{2}(I+Z_1Z_2)$, $B_x=\frac{1}{2}(I-X_1X_2)$ and $C_x=\frac{1}{2}(I+X_1X_2)$, where $X_i$ and $Z_i$ denotes Pauli matrices acting on the corresponding qubit $i$ ($i=1,2$). Starting with ``walkers" initially in an unpolarized state in any node, i.e. $\rho_0=\frac{1}{4}I_4\otimes|j\rangle\langle j|$, the OQW will converge to a state $\rho=\frac{1}{4}|\psi^-\rangle\langle\psi^-|\otimes|U,L\rangle\langle U,L|+\frac{1}{4}|\phi^-\rangle\langle\phi^-|\otimes|U,R\rangle\langle U,R|+\frac{1}{4}|\psi^+\rangle\langle\psi^+|\otimes|D,L\rangle\langle D,L|+\frac{1}{4}|\phi^+\rangle\langle\phi^+|\otimes|D,R\rangle\langle D,R|$. This means that measuring the position of the ``walkers" will determine corresponding Bell-state of their internal degrees of freedom \cite{note}.

\begin{figure}
\begin{center}
\includegraphics[width= 0.5\linewidth]{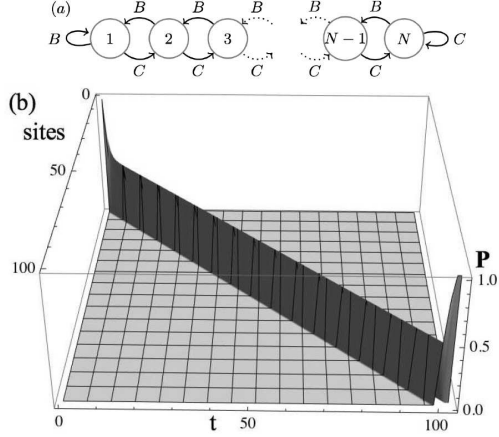}
\caption{Efficient transport with Open Quantum Walk. Fig. (a): a scheme of the chain of the $N$ nodes with neighbor-neighbor interaction. Fig. (b): occupation probability distribution as a function of time and lattice sites. The initial state of the walker is localized in the first node and given by $\rho_0=\frac{1}{2} I_2\otimes|1\rangle\langle 1|$. }
\end{center}
\end{figure}

Another application of the OQWs is a description of a dissipatively driven quantum bus between computational quantum registers. To this end we consider a chain of nodes (see Fig. 4(a)). Initially, the first node of the chain is in the exited state, so that $\rho_0=\frac{1}{2} I_2\otimes|1\rangle\langle 1|$. To be specific, we chose transition operators $B$ and $C$ as follows, $B=\sqrt{p}|\psi^{(1)}\rangle\langle\psi^{(2)}|$  and $C=\sqrt{q}|\psi^{(2)}\rangle\langle\psi^{(2)}|+|\psi^{(1)}\rangle\langle\psi^{(1)}|$. It is very interesting to note that in this case the state $|\psi^{(1)}\rangle$ propagates through the chain with velocity almost equal to 1 (in units of cells per time step): the initial excitation in node (1) is completely transferred to the last node (N) in N+2 steps. In Fig. 4 (b) we consider a 100 node chain with $\sqrt{p}=4/5$, $|\psi^{(1)}\rangle=|+\rangle$, $|\psi^{(2)}\rangle=|-\rangle$ and show that the initial excitation reaches the final node (100) in 102 steps. The high performance of transport of excitations in the OQW formalism opens up new avenues of research into the understanding of quantum efficiency in open systems.

OQWs include the discrete time version of the dissipative quantum computing (DQC) introduced by Verstraete \textit{et al.} \cite{Vers}. In the original setup a linear chain of time registers is considered and the initial state is prepared in the time register $0$. A quantum computation is performed by the dissipative evolution of the system into its unique steady state $\rho=\frac{1}{T+1}\sum_t|\psi_t\rangle\langle \psi_t|\otimes|t\rangle\langle t|$ \cite{Vers}. The result of the quantum computation can be read-out by measuring the state of the system in the last time register $T$ which is given by $|\psi_T\rangle=U_TU_{T-1}\ldots U_2U_1|\psi_0\rangle$, where $\{U_t\}^T_{t=1}$ is an appropriate sequence of unitary operators \cite{Vers}. The probability of a successful read-out is $1/(T+1)$. A discrete time version of DQC can be realized as an OQW on a linear chain of time registers (Fig. 5) by choosing the transition operators as it is shown in Fig. 5 and constants $\omega=\lambda=1/2$. However, with the same number of steps in the OQW formulation of DQC the probability of successful read-out can be increased arbitrarily close to one. In order to understand this dramatic improvement in efficiency we recall that in the original DQC formulation the probability of read-out of the final state is determined by the form of ``jumping" operators between  time registers, i.e., $L_t=U_t\otimes|t\rangle\langle t+1|+U_t^\dag\otimes|t+1\rangle\langle t|$. The probability to "jump" forward and backward in the time register is the same. In the OQW formulation of the DQC we have the freedom of choosing $\omega>\lambda$ which induces the steady state of the OQW with probability of read-out of the final state between $1/(T+1)$ and $1$.

\begin{figure}
\begin{center}
\includegraphics[width= 0.7\linewidth]{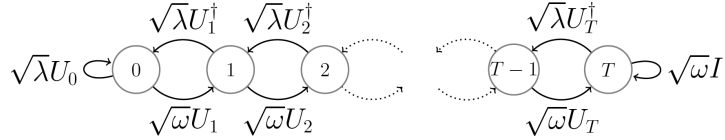}
\caption{OQW formulation of dissipative quantum computing (DQC). The initial state is prepared in the time-register $0$. After performing the OQW, the results of the algorithm can be readout from the time-register $N$. The internal state of the "walker" in the register $T$ will be given by $|\psi_T\rangle=U_T\ldots U_1|\psi_0\rangle$. The positive constants $\omega$ and $\lambda$ satisfy $\omega+\lambda=1$.}
\end{center}
\end{figure}

\section{Conclusion}

In conclusion, we have shown that the recently introduced formalism of OQWs is a very useful tool for the formulation of dissipative quantum computing algorithms and for  dissipative quantum state preparation. OQWs are to dissipative quantum computing what Hadamard quantum walks are to circuit based quantum computing. In particular, we have shown OQW implementation of circuit and dissipative models of quantum computing. Remarkably, the OQW discretisation of dissipative quantum computing increases the probability of successful implementation of the quantum algorithm with respect to the original formulation.  It is to be expected, that OQWs will lead to the optimal formulation of certain classes of quantum algorithms. Furthermore, we have indicated that OQW can be used  to explain non-trivial highly efficient transport phenomena not only in linear but also in more complex topologies of the underlying graphs. This implies that OQW as quantum walks which are driven by dissipation and decoherence are one of the candidates for understanding the remarkable transport efficiency in photosynthetic complexes \cite{qbt}. We expect the potential of this framework  to be soon revealed in the realms of quantum computing and quantum biology.

\section*{Acknowledgements}
This work is based upon research supported by the South African
Research Chair Initiative of the Department of Science and
Technology and National Research Foundation.
Work supported by ANR project ``HAM-MARK" N${}^\circ$ ANR-09-BLAN-0098-01


\bibliographystyle{elsarticle-num}
\bibliography{<your-bib-database>}

\begin{thebibliography}{00}
\bibitem{note} Please note that this is not a unique steady state of the system and it is depends on initial conditions. It is clear that this type of the random walk will not affect an initial state of the following form, $\rho=\alpha|\psi^-\rangle\langle\psi^-|\otimes|U,L\rangle\langle U,L|+\beta|\phi^-\rangle\langle\phi^-|\otimes|U,R\rangle\langle U,R|+\gamma|\psi^+\rangle\langle\psi^+|\otimes|D,L\rangle\langle D,L|+\delta|\phi^+\rangle\langle\phi^+|\otimes|D,R\rangle\langle D,R|$, where $\alpha, \beta, \gamma, \delta\ge0$ and $\alpha+\beta+\gamma+\delta=1$
\end{thebibliography}

\end{document}